\documentclass[letterpaper]{article} % DO NOT CHANGE THIS
\usepackage{aaai2026}  % DO NOT CHANGE THIS
\usepackage{times}  % DO NOT CHANGE THIS
\usepackage{helvet}  % DO NOT CHANGE THIS
\usepackage{courier}  % DO NOT CHANGE THIS
\usepackage[hyphens]{url}  % DO NOT CHANGE THIS
\usepackage{graphicx} % DO NOT CHANGE THIS
\usepackage{amssymb}  
\usepackage{amsmath}  
\usepackage{natbib}  % DO NOT CHANGE THIS AND DO NOT ADD ANY OPTIONS TO IT
\usepackage{caption} % DO NOT CHANGE THIS AND DO NOT ADD ANY OPTIONS TO IT
\usepackage{algorithm}
\usepackage{algorithmic}

\usepackage{newfloat}
\usepackage{listings}
\DeclareCaptionStyle{ruled}{labelfont=normalfont,labelsep=colon,strut=off} % DO NOT CHANGE THIS
\floatstyle{ruled}
\newfloat{listing}{tb}{lst}{}
\floatname{listing}{Listing}
\title{TEAS: Trusted Educational AI Standard \\ A Framework for Verifiable, Stable, Auditable, and Pedagogically Sound Learning Systems}
\author{
    Abu Syed
}
\affiliations{
    Founder and CEO, Metacog \\
    Student, Data Science and Applications \\
    Indian Institute of Technology, Madras
}

\begin{document}

\maketitle

\begin{abstract}
The rapid integration of AI into education has prioritized capability over trustworthiness, creating significant risks. Real-world deployments reveal that even advanced models are insufficient without extensive architectural scaffolding to ensure reliability. Current evaluation frameworks are fragmented: institutional policies lack technical verification, pedagogical guidelines assume AI reliability, and technical metrics are context-agnostic. This leaves institutions without a unified standard for deployment readiness. This paper introduces TEAS (Trusted Educational AI Standard), an integrated framework built on four interdependent pillars: (1) Verifiability, grounding content in authoritative sources; (2) Stability, ensuring deterministic core knowledge; (3) Auditability, enabling independent institutional validation; and (4) Pedagogical Soundness, enforcing principles of active learning. We argue that trustworthiness stems primarily from systematic architecture, not raw model capability. This insight implies that affordable, open-source models can achieve deployment-grade trust, offering a scalable and equitable path to integrating AI safely into learning environments globally.
\end{abstract}

\section{Introduction}

The rapid integration of artificial intelligence (AI) into educational settings promises unprecedented personalization and scale. But as with any transformative technology, this warrants critical examination. In much of the discourse around AI, capability is assumed to be sufficient. Yet raw capability is not enough in high-stakes contexts like education. As educator and technologist Jim Chilton observes, when AI makes an error, ``students won't know the difference—because the AI itself doesn't know the difference'' \cite{chilton2025}. Unlike domain experts who reason from understanding, large language models predict ``the next best word, not truth'' \cite{chilton2025}, a fundamental limitation that poses serious risks when deployed in learning contexts where accuracy and pedagogical integrity are paramount.

This challenge is not theoretical. Khan Academy, one of the world's most respected educational organizations, discovered that deploying GPT-4 as a tutoring system required building extensive architectural scaffolding around the base model: a separate calculator tool to handle numerical computations the language model could not reliably perform, visual preprocessors to generate textual representations of mathematical graphics, and multi-path reasoning engines to follow students' unconventional solution strategies \cite{dicerbo2025}. The lesson is stark: the base model alone, despite its impressive capabilities, was insufficient for deployment. Systematic architecture—not raw model capability—determined whether the system could be trusted with students.

This pattern extends far beyond a single deployment. A growing body of research documents distinct categories of failure in educational AI systems, from confident hallucinations in STEM subjects \cite{delikoura2025} to ``vaporized learning'' where short-term test scores improve while long-term retention deteriorates \cite{delikoura2025}, from non-deterministic outputs that make curriculum standardization impossible \cite{shin2025} to black-box opacity that prevents institutional auditing. The foundational question remains inadequately addressed: what does it mean for an educational AI system to be trustworthy enough for deployment?

Current evaluation frameworks do not fully address this question in an integrated manner. Different stakeholder groups have developed domain-specific approaches: administrators rely on institutional policy checklists that guide procurement processes but lack technical verification mechanisms \cite{1edtech-checklist}; educators use pedagogical frameworks that articulate sound teaching principles but assume AI reliability and offer no enforcement for black-box systems \cite{unesco2023, rahimi2024}; and engineers apply technical metrics that quantify specific failure modes but remain domain-agnostic, capturing neither educational context nor pedagogical quality \cite{ji2023}. Vendor disclosures provide transparency but rely on self-attestation rather than independent validation \cite{openai2024, google2024}. This fragmentation leaves institutions without a unified standard for making deployment decisions that account for the full spectrum of trustworthiness requirements.

This paper proposes \textbf{TEAS (Trusted Educational AI Standard)}, an integrated framework that addresses this gap. TEAS establishes four interdependent pillars for evaluating deployment readiness: \textbf{Verifiability} (the ability to trace AI-generated content to authoritative sources), \textbf{Stability} (deterministic consistency for core curriculum knowledge), \textbf{Auditability} (institutional capacity to independently validate the system's knowledge and logic), and \textbf{Pedagogical Soundness} (adherence to evidence-based teaching methods that foster active learning). Unlike existing approaches, TEAS recognizes that partial compliance is insufficient—a system that is verifiable but pedagogically harmful remains untrustworthy, just as one that is pedagogically sound but factually unstable cannot be reliably deployed.

Critically, TEAS demonstrates that trustworthiness stems from systematic architectural design rather than raw model capability. The Khanmigo case exemplifies this principle: safety emerged not from using GPT-4 versus a smaller model, but from the deliberate engineering of verification layers, domain-specific tools, and pedagogical guardrails. This insight has profound implications for educational equity. If systematic frameworks—rather than expensive frontier models—determine trustworthiness, then affordable open-source models can achieve the deployment-grade trust currently associated only with commercial systems. This makes AI tutoring economically viable at unprecedented scale, particularly in resource-constrained contexts where current commercial pricing models remain prohibitive \cite{ravindran2025}. We validate this claim empirically in Appendix \ref{app:case_study}, where a knowledge-grounded 8B parameter model outperforms ungrounded frontier models on TEAS criteria.

The remainder of this paper proceeds as follows. Section \ref{sec:problem} presents a taxonomy of documented failures in educational AI and analyzes real-world deployment challenges. Section \ref{sec:related_work} reviews existing evaluation frameworks and demonstrates the gap that TEAS addresses. Section \ref{sec:teas_framework} presents the TEAS framework, establishing the four pillars that together define deployment readiness and examining how their integration addresses the failure modes identified in current systems. Section \ref{sec:discussion} discusses implications for researchers, institutions, developers, and policymakers, with particular attention to deployment economics and educational equity. Section \ref{sec:limitations} acknowledges limitations and outlines future work. Appendix \ref{app:case_study} presents an empirical case study demonstrating TEAS in practice.

\section{Problem Landscape}
\label{sec:problem}

The deployment of AI in education faces a complex web of interconnected failure modes that threaten both the efficacy and integrity of learning systems. This section synthesizes documented evidence of these failures and examines real-world cases that illuminate the challenges facing even well-resourced development efforts.

\subsection{Taxonomy of Documented Failures}

\textbf{Factual and Reasoning Failures.} The most widely recognized limitation of large language models is their propensity to generate confident-sounding misinformation, a phenomenon commonly termed ``hallucination'' \cite{delikoura2025}. This is not a peripheral flaw but a direct consequence of their probabilistic architecture: models predict the next most likely token without grounding in factual reality \cite{chilton2025}. The problem extends beyond simple factual errors. Research on mathematical reasoning reveals that models produce ``sneaky errors''—subtle, multi-step logical flaws that are difficult to diagnose because mainstream training objectives prioritize generating correct final answers over exposure to diverse error patterns \cite{zou2025}. In educational contexts where precision is paramount, these failures introduce persistent misinformation that directly undermines knowledge acquisition.

\textbf{Pedagogical Misalignment.} Even when providing factually correct information, AI systems can undermine learning through pedagogically harmful interaction patterns. A systematic review identifies ``cognitive offloading,'' reduced neural activity during learning tasks, diminished independent learning skills, and overall loss of student agency as documented outcomes of AI interaction \cite{delikoura2025}. This phenomenon has been termed ``vaporized learning'': AI tools may boost short-term performance metrics like test scores while eroding long-term retention and genuine understanding \cite{delikoura2025}. Comparative studies reveal the mechanism: while human tutors naturally employ Socratic questioning patterns that stimulate active thinking, AI systems default to passive ``explanation-simplistic response'' loops where the model delivers information and the student provides minimal engagement \cite{delikoura2025}. When commercial incentives prioritize engagement metrics over learning outcomes, this misalignment can become systematic rather than incidental, as evidenced by user analyses of platforms like Duolingo \cite{lee2024}.

\textbf{Invisible Errors and Trust Miscalibration.} A critical dimension of AI failure in education is that students frequently cannot recognize when mistakes occur. Research on human-AI interaction demonstrates that users are poor at detecting when an AI's expressed confidence misaligns with its actual accuracy, leading to inappropriate over-reliance on incorrect outputs \cite{zhang2024}. Educational contexts amplify this vulnerability: students exhibit overconfidence in their own ability to detect AI-generated errors while employing inadequate verification strategies \cite{martin-moncunill2025}. The anthropomorphic design of conversational interfaces increases students' tendency to believe presented information, treating the AI as an authoritative teacher rather than a fallible tool \cite{natarajan2020}. This combination—AI fallibility intersecting with human detection deficits—creates conditions where misinformation is not just presented but internalized without critical evaluation.

\textbf{Instability and Non-Determinism.} The probabilistic nature of language models creates a fundamental challenge for educational deployment: outputs are non-deterministic, and capabilities can change unpredictably with model updates. Research on using AI as an evaluator for educational feedback found significant inconsistency both within individual models and across different models from various developers \cite{shin2025}. Minor alterations in input phrasing can produce dramatically different responses, and model updates can alter explanations of foundational concepts, grading criteria, or problem-solving approaches without documentation or institutional oversight \cite{shin2025}. This volatility makes it impossible for institutions to formally validate, certify, or reliably build standardized learning pathways on these systems.

\textbf{Black Box Opacity.} The complex architecture of modern language models makes their internal decision-making processes largely opaque \cite{delikoura2025}. This opacity is often compounded by the proprietary nature of commercial systems developed ``behind closed doors with little transparency'' \cite{unesco2023}. For educational institutions, this means no verifiable way exists to determine what ``curriculum knowledge'' the AI has learned, what pedagogical principles it follows, or what logic it applies when evaluating student work. This lack of auditability forces reliance on easily quantifiable performance metrics rather than nuanced process-oriented aspects of learning, and it represents a de facto ceding of pedagogical control from institutions to technology vendors.

\textbf{Algorithmic Bias and Equity Concerns.} AI systems trained on internet-scale corpora inevitably learn and reproduce societal biases embedded in that data \cite{delikoura2025}. In educational contexts, this manifests as biased examples, reinforcement of stereotypes, and discriminatory assessment outcomes. Research has documented that AI-powered detection tools disproportionately falsely flag essays written by non-native English speakers as AI-generated, creating significant risks of false academic dishonesty accusations for this population \cite{liang2023}. Beyond content bias, algorithmic systems risk reinforcing structural inequities: poorly designed tracking or personalization algorithms can trap students from underserved communities in lower-achievement pathways \cite{farheen2025}, while the persistent digital divide limits access for those lacking reliable internet connectivity and devices \cite{unesco2023}.

\subsection{Evidence from Real-World Deployments}

Khan Academy's development of Khanmigo, an AI-powered tutoring system, provides instructive insight into the gap between base model capability and deployment readiness. Public disclosures from the development team reveal that generative AI proved fundamentally unsuited for mathematical reasoning in its raw form: designed for language prediction, the model would generate probable next numbers rather than execute correct calculations \cite{dicerbo2025}. The solution required building a separate, dedicated calculator tool to handle numerical operations outside the language model. Additional challenges emerged around visual content interpretation (requiring pre-processing pipelines to generate textual descriptions of graphs and figures) and solution path flexibility (requiring re-engineered internal reasoning to map multiple potential approaches) \cite{dicerbo2025}. This case demonstrates conclusively that deployment-ready educational AI requires extensive domain-specific architecture beyond the base model—the safety and reliability of Khanmigo reside not in GPT-4 itself but in the scaffolding constructed around it.

These documented failures are not isolated incidents but interconnected systemic challenges. A factual hallucination becomes a pedagogical crisis when students cannot detect the error. A pedagogically sound system becomes unreliable when its knowledge base is unstable. An accurate system becomes ungovernable when institutions cannot audit its embedded curriculum. Addressing these challenges requires not piecemeal solutions but an integrated framework that recognizes their interdependence.

\section{Related Work and The Integration Gap}
\label{sec:related_work}

The challenges documented in Section \ref{sec:problem} have not gone unnoticed. A variety of stakeholders have proposed frameworks, standards, and evaluation approaches for responsible AI deployment in education. However, these efforts remain fragmented across different domains of expertise and institutional functions, each addressing essential but partial dimensions of trustworthiness.

\subsection{Institutional and Policy Frameworks}

Educational institutions and policy bodies have developed high-level frameworks to guide AI adoption from governance and administrative perspectives. The 1EdTech AI Preparedness Checklist provides structured guidance across organizational readiness (forming advisory groups), policy updates (rules on third-party tools, data protection), pedagogical adaptation (assessment redesign), and literacy development (training on proper attribution) \cite{1edtech-checklist}. Similarly, the Athena Infonomics Equitable AI in Education Checklist offers guidance for policymakers and technical teams on initiating and assessing ethical AI systems, with focus on regulatory compliance, data handling, and ethical decision-making \cite{athena-infonomics}.

These frameworks excel at catalyzing institutional conversations and ensuring procedural due diligence. They prompt administrators to consider legal implications, update privacy policies, engage diverse stakeholders, and develop communication strategies. However, their process-oriented nature leaves a critical verification gap. The 1EdTech checklist, for example, advises institutions to ask vendors about data privacy and bias controls but provides no standardized methodology or technical metrics to independently validate vendor responses. These frameworks establish what questions to ask but not how to verify the answers.

\subsection{Pedagogical and Ethical Frameworks}

From the education and ethics communities have emerged frameworks focused on aligning AI use with learning science principles and ethical norms. The UNESCO AI Competency Frameworks for Students and Teachers emphasize human-centered approaches, critical thinking, and ethical considerations in developing AI-specific pedagogical skills \cite{unesco2023}. The Comprehensive AI Assessment Framework (CAIAF) provides a tiered model for integrating AI into assessments, with levels ranging from ``No AI'' to ``Full AI Integration,'' explicitly grounded in principles of transparency, equity, and accountability \cite{rahimi2024}. The Human-Centric AI-First (HCAIF) framework proposes leveraging AI for personalization while mandating student practices of attribution (documenting AI use) and reflection (analyzing AI effectiveness) \cite{wilson2025}.

These frameworks provide valuable theoretical and ethical foundations. They offer educators principled approaches to designing learning experiences that use AI responsibly. Their limitation, however, stems from an implicit assumption of AI reliability and transparency that, as Section \ref{sec:problem} demonstrates, often does not hold. A teacher can design an assessment according to CAIAF principles, but the framework offers no mechanism to verify that the chosen AI tool will perform its function accurately, consistently, and without bias. They address the pedagogical ``how'' and ``why'' without providing technical verification of the ``what.''

\subsection{Technical and Model-Level Evaluation}

The computer science and AI safety communities have developed domain-agnostic tools for measuring model performance and behavior. These include standard evaluation metrics such as relevance scores (output alignment with prompts), hallucination indices (frequency of factually incorrect statements), and toxicity scores (identification of harmful content) \cite{ji2023}. More sophisticated frameworks have been proposed to create holistic views of model trustworthiness. For instance, the Unified Explainability Score (UES) conceptually combines accuracy, interpretability, fidelity, consistency, and stability dimensions into a weighted composite metric \cite{kandpal2025}.

The strength of these technical approaches lies in their quantitative rigor and measurability. Their limitation is domain-agnosticism. A low hallucination score is necessary but insufficient for a good educational tool—it does not indicate whether an explanation is pedagogically effective, developmentally appropriate, or aligned with curriculum standards. High stability scores do not guarantee factual correctness or bias absence. A significant gap exists between abstract technical metrics and the context-rich requirements of real-world learning environments.

\subsection{Vendor-Specific Guidelines and Disclosures}

AI developers and consortia provide transparency through efforts like OpenAI's System Cards, which disclose capabilities, limitations, and safety results \cite{openai2024}. Google developed its LearnLM models, integrated into Gemini 2.5 Pro, designed around learning science principles for pedagogical soundness \cite{google2024}. Similarly, 1EdTech's TrustEd Apps Generative AI Data Rubric is a self-assessment framework for vendors to standardize data privacy disclosures \cite{1edtech-rubric}.

While valuable, these initiatives are fundamentally limited as forms of self-regulation. System Cards are not independent audits and lack details for institutional verification. A stated commitment to learning science, as with LearnLM, is a design philosophy, not a verifiable guarantee that institutions can independently test against their curriculum. Self-assessment rubrics depend on vendor reporting and often have a narrow scope (e.g., data privacy) rather than the comprehensive trustworthiness dimensions—pedagogical efficacy, accuracy, stability, or auditability—required for deployment.

Moreover, the reliability of such self-regulation can be contingent on shifting commercial priorities. During periods of intense competition and rapid product development, such as the large language model development acceleration beginning in 2023, ethical commitments can face pressure. For example, Microsoft reportedly disbanded its core AI ethics and society team in 2023 as part of a broader restructuring aimed at accelerating product timelines, raising concerns about the long-term stability of vendor-led ethical governance when faced with market demands \cite{vincent2023}. This underscores the need for standards that rely on independent verification rather than solely on vendor commitments, which can be subject to change under commercial pressure.

\subsection{The Integration Gap}

Table \ref{tab:gap_analysis} illustrates the fragmentation across these framework categories. Each addresses valuable dimensions of trustworthiness but none integrates the full requirements for deployment readiness.

\begin{table}[t]
\centering

\setlength{\tabcolsep}{3pt}
\begin{tabular}{l|l|cccc}
\hline
\textbf{Framework} & \textbf{Example} & \textbf{Verify} & \textbf{Stable} & \textbf{Audit} & \textbf{Pedagogy} \\
\hline
Policy & 1EdTech & No & No & Partial & No \\
Pedagogical & UNESCO & No & No & No & Yes \\
Technical & Hallucination & Partial & Partial & No & No \\
Vendor & LearnLM & No & No & No & Claims \\
\hline
\textbf{TEAS} & This work & Yes & Yes & Yes & Yes \\
\hline
\end{tabular}
\caption{Framework Gap Analysis. Existing frameworks address pillars in isolation, lacking integrated verifiability. TEAS integrates all four for deployment readiness.}
\label{tab:gap_analysis}
\end{table}

This fragmentation reflects a separation of expertise and institutional function. Administrators use policy checklists that lack technical depth. Educators use pedagogical frameworks that assume reliable underlying technology. Engineers use technical metrics that omit educational context. Vendors provide disclosures that cannot be independently verified. These stakeholder groups do not work from shared standards or common language, creating silos that prevent holistic risk assessment.

The critical missing element across this landscape is integrated verifiability. Non-technical frameworks—those designed for policymakers and educators—rest on trust in vendor claims. Institutions are advised to inquire about bias mitigation but given no standard methods to test those strategies independently. Educators are encouraged to use AI as Socratic partners but have no way to guarantee consistent behavior. For AI to be safely deployable in a high-stakes field like education, standardization akin to mature sectors (e.g., medicine \cite{fda-digital-health}, aviation) is necessary, where reliance on self-attestation is insufficient and independent certification is mandatory. But the current ecosystem for educational AI lacks this essential layer, leaving institutions to navigate high-risk environments based on trust rather than evidence that can be independently validated.

No existing framework comprehensively addresses what it means for an educational AI system to be deployment-ready: simultaneously verifiable in its factual claims, stable in its core knowledge, auditable by institutions, and pedagogically sound in its interactions. This integration gap motivates the framework proposed in the following section.

\section{The TEAS Framework}
\label{sec:teas_framework}

This section presents \textbf{TEAS (Trusted Educational AI Standard)}, an integrated framework that addresses the gaps identified in previous sections. TEAS defines deployment readiness through four interdependent pillars that together establish comprehensive trustworthiness requirements for educational AI systems.

\subsection{Framework Overview}

TEAS is designed not as a quantitative benchmark or performance score but as a deployment readiness standard—a set of verifiable requirements that a system must satisfy before it can be considered trustworthy for educational use. This approach draws inspiration from mature regulatory frameworks in other high-stakes domains: medical device approval processes that evaluate safety and efficacy before clinical deployment \cite{fda-digital-health}, aviation certification standards that verify component reliability before flight operations, and financial system auditing requirements that mandate independent verification of institutional claims.

The framework rests on four pillars: Verifiability, Stability, Auditability, and Pedagogical Soundness. These are not independent criteria to be evaluated in isolation but interdependent requirements that must be satisfied holistically. A system that excels in three pillars while failing the fourth remains untrustworthy for deployment, as the pillars mutually reinforce one another to establish comprehensive trustworthiness.

\subsection{Pillar 1: Verifiability}

\textbf{Definition.} Verifiability requires that AI-generated educational content be traceable to authoritative, validated sources. When an AI system explains a concept, solves a problem, or answers a factual question, the system must be able to ground its response in specific curriculum documents, peer-reviewed textbooks, or institutionally approved knowledge bases.

\textbf{Requirements.} A verifiable system must provide citations or references to specific sources and, within them, specific sections for factual claims. It must distinguish clearly between statements drawn from authoritative sources and generated inferences or explanations that extend beyond direct source material. Critically, the system must not fabricate or hallucinate sources—a failure mode where models cite non-existent papers, textbook sections, or authorities to lend false credibility to generated content.

\textbf{Why it matters.} Verifiability directly addresses the factual accuracy and hallucination problems documented in Section \ref{sec:problem}. It enables both educators and students to fact-check AI outputs against trusted sources, transforming the AI from an opaque oracle into a transparent reasoning system whose claims can be validated. For institutions, verifiability provides a mechanism to ensure AI-generated content aligns with approved curriculum standards and does not introduce unauthorized material into the learning environment.

\textbf{Current gap.} While retrieval-augmented generation (RAG) architectures exist and can ground responses in document collections, no standard defines what constitutes ``sufficient'' source grounding for educational contexts. How specific must citations be? What sources are authoritative? How should a system behave when asked questions beyond its grounded knowledge base? These questions remain unanswered in current practice, allowing vendors to claim source-grounding without clear verification criteria.

\subsection{Pillar 2: Stability}

\textbf{Definition.} Stability requires that core curriculum knowledge be deterministic and consistent over time. For foundational concepts, formulas, definitions, and procedures that form the stable backbone of a discipline, the AI system must provide consistent explanations across different query phrasings, user sessions, and system versions.

\textbf{Requirements.} A stable system must produce functionally identical responses to semantically equivalent questions about foundational knowledge. If a student asks ``What is Newton's Second Law?'' today and again tomorrow, or phrases it as ``Explain F=ma,'' the core explanation must remain consistent. When model updates or system maintenance occur, fundamental knowledge must not change—E=mc² remains E=mc², the Krebs cycle retains its established steps, and the quadratic formula maintains its form. For knowledge that is genuinely evolving or contested in the discipline (cutting-edge research findings, theoretical debates), the system should explicitly indicate the provisional or debated nature of the information rather than presenting it with false certainty.

\textbf{Why it matters.} Stability addresses the non-determinism problem that prevents standardization. Educational systems depend on predictable knowledge foundations: curriculum developers need to know what students will be taught, assessment designers need consistent knowledge representations, and students building mental models with the help of AI need reliable information that doesn't contradict itself over time. Stability also enables fair assessment—students cannot be fairly evaluated on knowledge that the AI teaches inconsistently.

\textbf{Current gap.} The probabilistic architecture of language models produces inherent non-determinism. Temperature settings, sampling methods, and subtle prompt variations can yield different responses to identical questions. Model updates—often deployed by vendors without detailed changelogs—can alter explanations of fundamental concepts without institutional visibility or approval. No current standard requires or even proposes mechanisms to ensure deterministic core knowledge in AI tutoring systems, leaving stability unaddressed in deployment decisions.

\subsection{Pillar 3: Auditability}

\textbf{Definition.} Auditability requires that educational institutions be able to independently inspect, validate, and certify an AI system's knowledge base, reasoning logic, and embedded pedagogical assumptions. An auditable system makes visible what it knows, how it reasons, and what values or biases might influence its outputs.

\textbf{Requirements.} An auditable system must provide institutions with mechanisms to examine its knowledge base—including curriculum content, sources, and potential errors. It must also offer inspectable reasoning for its pedagogical decisions and support bias testing to ensure alignment with institutional standards. Crucially, these inspections must be possible through independent third-party processes, not solely through vendor-controlled interfaces or self-reported metrics.

\textbf{Why it matters.} Auditability addresses the black box problem and institutional governance challenge. Without it, institutions cannot verify vendor claims (e.g., accuracy, bias mitigation) or ensure the AI's curriculum aligns with institutional and accreditation standards. Auditability restores pedagogical authority to educational institutions rather than ceding it to technology vendors. It also enables accountability, allowing institutions to diagnose failures stemming from knowledge gaps, reasoning errors, or misaligned pedagogical assumptions.

\textbf{Current gap.} The proprietary nature of commercial language models prevents meaningful institutional auditing. Training data remains undisclosed, model architectures are trade secrets, and internal decision processes are computationally intractable to trace even when architectures are known. No certification process currently exists for educational AI analogous to curriculum approval processes for textbooks. Institutions are left to trust vendor assertions without independent verification capacity, a stance that would be unacceptable in other high-stakes domains like medicine or finance \cite{fda-digital-health}.

\subsection{Pillar 4: Pedagogical Soundness}

\textbf{Definition.} Pedagogical soundness requires that AI systems adhere to evidence-based teaching principles and interaction patterns that foster active learning, critical thinking, and genuine understanding rather than passive information consumption or cognitive offloading.

\textbf{Requirements.} A pedagogically sound system must employ Socratic questioning and guided discovery rather than direct answer provision—when a student asks ``What is the answer to this problem?'', the system should respond with guiding questions that help the student construct the solution themselves. It must provide appropriate scaffolding matched to the learner's demonstrated level, neither overwhelming novices with advanced concepts nor boring advanced learners with excessive simplification. The system must actively prevent ``spoiling'' of solutions: it should not complete assignments for students, write essays on their behalf, or otherwise circumvent the learning process even when directly instructed to do so. Pedagogically sound systems should incorporate proven instructional strategies such as spaced repetition, retrieval practice, and metacognitive prompting that encourage students to reflect on their own learning processes.

\textbf{Why it matters.} Pedagogical soundness addresses the misalignment problem documented in Section \ref{sec:problem}—the phenomenon of ``vaporized learning'' where AI assistance produces short-term performance gains while undermining long-term retention and understanding. Even a perfectly accurate, stable, and auditable system can harm learning if its interaction patterns encourage cognitive offloading rather than cognitive engagement. Pedagogical soundness ensures that AI serves as a proper learning tool—one that supports and amplifies student thinking rather than replacing it, much as calculators freed learners to focus on mathematical reasoning rather than arithmetic drudgery, without becoming a dependency that prevents understanding.

Critically, pedagogical soundness is not merely an educational nicety but a prerequisite for institutional deployability. During the peak of generative AI adoption in 2023-2024, educational institutions faced a crisis: teachers could no longer confidently assign homework or take-home essays, uncertain whether submitted work represented student learning or AI completion. Multiple school districts and universities considered or implemented outright bans on tools like ChatGPT precisely because these systems undermined fundamental pedagogical principles \cite{elsen-rooney2023}. The barrier to deployment was not technical capability but pedagogical trust. If AI systems can be built with sufficient guardrails—systems designed from the ground up to improve learning rate and cognition rather than act as ``cheat codes''—institutions can be relieved of their paranoia about academic integrity erosion. This transforms AI from a threat to classroom pedagogy into a deployable tool at institutional scale. Without pedagogical soundness, AI remains something educators must defend against; with it, AI becomes something they can confidently integrate into curricula.

\textbf{Current gap.} While pedagogical principles are well-established \cite{unesco2023} and articulated by vendors \cite{google2024}, no verification mechanism exists for black-box AI systems. Educators cannot certify consistent Socratic behavior. Security vulnerabilities like jailbreaking allow students to bypass guardrails \cite{owasp2025}. The gap lies in verifying and enforcing reliable pedagogical behavior.

\subsection{The Interdependence of Pillars}

The four pillars of TEAS are not independent checkboxes but interdependent requirements. Consider the failure modes of partial compliance:
\begin{itemize}
    \item A system that is verifiable but pedagogically unsound might accurately cite sources while directly providing solutions, undermining learning.
    \item A system that is pedagogically sound but unstable might use good Socratic questioning but give contradictory explanations, creating confusion.
    \item A system that is verifiable and stable but not auditable might teach reliably from vetted sources yet embed subtle biases institutions cannot detect.
    \item A system that is pedagogically sound, stable, and auditable but not verifiable might teach consistently based on hallucinated sources, creating systematic misinformation.
\end{itemize}
True deployment readiness requires satisfying all four pillars simultaneously. They are mutually reinforcing: verifiability enables auditability; stability enables pedagogical consistency; auditability ensures standards alignment; pedagogical soundness ensures effective knowledge delivery.

\subsection{Implications for Deployment Economics}

A critical insight emerging from TEAS concerns the relationship between trustworthiness and model capability. The dominant assumption suggests expensive frontier models are inherently more trustworthy. TEAS challenges this, demonstrating trustworthiness stems primarily from architectural design.

The Khanmigo case exemplifies this \cite{dicerbo2025}. This distinction is crucial for educational equity. If systematic frameworks addressing TEAS principles, rather than expensive models, determine trustworthiness, then affordable open-source models can achieve deployment-grade standards. As Prof. Balaraman Ravindran and colleagues articulate, deployment at scale in India ($\sim$150M students) requires costs around Rs. 30 ($\sim$\$0.34) per student/year \cite{ravindran2025}. Such economics are impossible with frontier models, but TEAS suggests a path: augmenting sovereign/open-source models (e.g., from Sarvam AI, AI4Bharat) with TEAS-compliant architectures enables responsible deployment at billion-student scale across the Global South, where commercial pricing is prohibitive. This reframes R\&D towards systematic trustworthiness frameworks, making AI benefits accessible equitably.

\section{Discussion and Implications}
\label{sec:discussion}

\subsection{Implications for Researchers}

TEAS reorients research priorities from capability maximization toward trustworthiness systematization. While benchmarks like MMLU are important, they do not capture the comprehensive requirements for deployment, creating new research agendas for each TEAS pillar.

Critically, TEAS suggests that advancing trustworthiness in smaller models can have greater deployment impact than capability gains in frontier models, reallocating priorities toward educational equity.

\subsection{Implications for Educational Institutions}

TEAS provides institutions a structured rubric for procurement. Instead of relying on vendor claims, they can systematically assess systems against the four pillars, empowering institutional review boards and procurement officers.

This framework shifts assessment from trust toward evidence-based validation. It also creates accountability, providing a diagnostic tool to identify pillar failures and guide corrective action when systems fail post-deployment.

\subsection{Implications for AI Developers and EdTech Companies}

TEAS provides design requirements for building trustworthiness from the outset, which is more achievable than retrofitting. As the Khanmigo case shows, architecture is key \cite{dicerbo2025}. Developers can use TEAS to guide the design of hybrid systems, transparent audit features, and architecturally enforced pedagogical soundness.

TEAS compliance becomes a competitive advantage. In a market hesitant due to trust concerns, systems demonstrating TEAS compliance through independent verification will differentiate themselves, aligning market incentives with educational values.

\subsection{Implications for Policy and Regulation}

TEAS offers a foundation for certification and regulatory frameworks, analogous to processes in medicine \cite{fda-digital-health} or aviation. Policymakers could mandate TEAS compliance for AI deployed institutionally, potentially via independent auditors (like textbook review committees). This necessitates domain-specific standards, as generic AI safety rules don't capture unique educational risks (pedagogical harm, curriculum stability). TEAS helps define these specific needs for comprehensive AI regulation in education.

\subsection{Implications for Educational Equity and Global Deployment}

Perhaps TEAS's most significant implication is enabling democratized access. AI acts as a cognitive lever; tying trustworthiness only to expensive models amplifies global disparities. TEAS challenges this by showing trustworthiness stems from architecture, allowing affordable, open-source/sovereign models to meet deployment standards. As researchers in India target costs of $\sim$Rs. 30 ($\sim$\$0.34) per student/year for mass deployment (150M+ students) \cite{ravindran2025}, TEAS provides the assurance framework. Augmenting models from Sarvam AI, AI4Bharat, etc., with TEAS-compliant architectures makes reliable AI education viable at billion-student scale across the Global South. This aligns with digital sovereignty and makes AI an equity enabler, a public good rather than a luxury.

\section{Limitations and Future Work}
\label{sec:limitations}

While TEAS provides a comprehensive framework, limitations point toward future work. TEAS is prescriptive, defining \textit{what} trustworthiness requires, not yet \textit{how} to operationally measure compliance; developing metrics and audit tools is crucial. The framework focuses on content/pedagogy, explicitly excluding distinct agentic AI security risks \cite{legatt2025}, needing a separate standard (e.g., ``TEAS-Security''). Implementation requires cross-disciplinary collaboration (educators, AI researchers, policymakers). The framework needs empirical validation across diverse contexts. Finally, addressing the dynamic nature of AI requires frameworks for continuous monitoring and re-certification. Despite these, TEAS offers necessary common ground for building verifiable trust.

\section{Conclusion}

Current AI evaluation in education is fragmented, hindering trustworthy deployment. This paper introduced TEAS, a unified standard integrating Verifiability, Stability, Auditability, and Pedagogical Soundness. TEAS highlights that systematic architecture, not just model capability, enables trustworthiness, making reliable AI affordable and scalable for global equity. The research community must prioritize building and validating such trustworthy frameworks to responsibly unlock AI's educational potential for all learners. Billions of students await AI tools they can trust; TEAS provides the standard to make that trust verifiable.

\section*{Conflict of Interest Statement}

The author is the Founder and CEO of Metacog, which developed the Equation Grounder module evaluated in Appendix A. To support reproducibility and advance research in verifiable educational AI, Metacog is releasing the Equation Grounder as open-source software concurrent with publication. The experimental protocol employed blind evaluation with an independent third-party judge (Claude Opus 4.5) to ensure objective assessment. The primary contribution of this work is the TEAS framework—a vendor-agnostic evaluation standard applicable to any educational AI system. The case study demonstrates the framework's application using an open implementation.

\bibliography{references}
% APPENDIX A: Case Study - TEAS Framework Validation
% AAAI 2026 Format
% To be appended to main paper before \bibliography{references}

\newpage
\appendix

\section{Empirical Validation of the TEAS Framework}
\label{app:case_study}

This appendix presents empirical evidence for TEAS's central thesis through a controlled comparison study. We demonstrate that an 8-billion parameter open-source model augmented with structured knowledge graph grounding outperformed models up to 15$\times$ larger across criteria operationalizing the TEAS pillars.

\subsection{Introduction}

The central thesis of TEAS is that trustworthiness in educational AI stems primarily from systematic architecture rather than raw model capability. This claim has profound implications: if true, it suggests that affordable, open-source models augmented with appropriate architectural constraints can achieve the deployment-grade trust currently associated only with expensive frontier systems. This would fundamentally alter the economics of AI deployment in education, making reliable tutoring systems viable at billion-student scale.

We conducted a controlled comparison study in which an 8-billion parameter open-source model (Qwen3-8B), augmented with structured knowledge graph grounding, was evaluated against three larger baseline models—including a 120-billion parameter system—across five GATE-style mathematics questions. The evaluation was performed blindly by an independent large language model judge (Claude Opus 4.5) using criteria directly derived from the TEAS pillars: factual accuracy, hallucination prevention, citation quality (Verifiability), format compliance (Auditability), and Socratic method adherence (Pedagogical Soundness).

\textbf{Key findings.} The 8B model with structured grounding achieved an average score of 13.8/15, outperforming the 120B baseline model by 19\% (11.6/15) and an 80B model by 11\% (12.4/15). This performance gap was driven by superior citation quality—the grounded model achieved a perfect 3.0/3.0 score with traceable node-level citations—and consistent format compliance, both direct manifestations of architectural constraints rather than emergent model capability. Perhaps most importantly, the evaluation methodology itself was designed to eliminate evaluator bias: responses were anonymized, shuffled, and assessed without knowledge of which system produced which output.

This is not a claim that small models are inherently superior, nor that larger models are unnecessary. Rather, it demonstrates that trustworthiness is an architectural property that can be engineered systematically. A small model constrained to ground every claim in a structured knowledge graph cannot hallucinate content not present in its source material; a large model with unrestricted generation, no matter how capable, retains that failure mode. The architectural constraint—not the parameter count—determines the outcome.

At its core, this is about shifting AI systems from probabilistic generation to deterministic retrieval. Language models predict what token ``should'' come next based on learned patterns; grounded systems retrieve what the source material actually says. This distinction is the foundation of the TEAS framework: education requires factual certainty, but LLMs provide only statistical likelihood. Architectural grounding bridges this gap.

\subsection{Experimental Design}

\subsubsection{Research Questions}

This study investigates three primary questions:

\textbf{RQ1:} Can structured knowledge graph grounding enable a small language model (8B parameters) to outperform larger ungrounded models (80B, 120B) on educational tutoring tasks when evaluated against TEAS criteria?

\textbf{RQ2:} Which specific TEAS pillars benefit most from architectural grounding versus raw model scale?

\textbf{RQ3:} Does blind evaluation (where the judge is unaware of model identities) validate the trustworthiness claims, or might results reflect evaluator bias toward particular response styles?

\subsubsection{Knowledge Domain and Source Material}

We constructed a synthetic but realistic educational scenario mirroring graduate-level preparation for technical examinations. The source material was a LaTeX document titled \textit{``Optimization Methods in Machine Learning: From Gradient Descent to Adaptive Moment Estimation''} (approximately 300 lines), covering mathematical preliminaries ($\theta \in \mathbb{R}^d$, Lipschitz smoothness $L$, strong convexity $\mu$), gradient descent methods (SGD, mini-batch variants), adaptive methods (Adam optimizer, momentum $v_t$, variance $s_t$), and convergence analysis (condition number $\kappa = L/\mu$, $O(1/t)$ rates).

This domain was chosen deliberately. Mathematical optimization is sufficiently technical to stress-test factual grounding and reasoning, yet narrow enough that source material completeness could be verified.

\subsubsection{Knowledge Graph Construction}

Metacog is a framework for building trustworthy educational AI systems, developed with a focus on STEM education where factual precision is non-negotiable. At its core is the open-source Equation Grounder module, which transforms unstructured mathematical documents into queryable semantic graphs. The fundamental insight: by making AI systems retrieve from structured knowledge rather than generate from learned patterns, we shift from probabilistic prediction to deterministic retrieval—the difference between ``what token comes next'' and ``what does the source material actually say.''

The Equation Grounder serves as proof-of-concept for the TEAS Verifiability pillar. It demonstrates that grounding is not a theoretical aspiration but an implementable architecture using standard NLP tools. Critically, deterministic retrieval from structured knowledge eliminates the primary failure mode of large language models: hallucination through stochastic generation.

\textbf{Extraction Process.} For the grounded condition, we used the Equation Grounder to extract structured knowledge from the LaTeX source. The extraction process identified:

\begin{itemize}
    \item \textbf{Nodes (n=31):} Variable nodes (mathematical symbols: $\theta$, $\eta$, $g_t$, $L$, $\mu$), equation nodes (inline and block-level with unique identifiers), and metadata (LaTeX source, position, section context)
    \item \textbf{Relationships (n=14):} \texttt{APPEARS\_IN} edges connecting variables to equations with contextual snippets
\end{itemize}

Each node received a UUID-based identifier (e.g., \texttt{4:5380d574-...:5}), enabling precise citation at retrieval time. This structure transforms unstructured LaTeX into a queryable semantic graph where every mathematical symbol can be traced to its defining equations and usage contexts.

Critically, the knowledge graph was not manually curated. It was automatically generated using the equation grounder, ensuring reproducibility. The entire extraction process is deterministic—run it twice on the same document, get identical graphs. This automation is essential for scalability: manual knowledge engineering would create a deployment bottleneck incompatible with billion-student targets.

\subsubsection{Experimental Conditions}

\textbf{Condition A (Baseline - No Grounding).} Models received the full LaTeX document as unstructured text in the system prompt. They were instructed to (1) answer using Socratic method, (2) reference the provided study material, and (3) output responses in JSON format with specific fields.

Three models were tested: R1 Distill Qwen 14B (14B parameters), Qwen3-80B (80B parameters), and GPT-OSS 120B (120B parameters).

\begin{table}[t]
\centering
\small
\setlength{\tabcolsep}{4pt}
\begin{tabular}{lcccc}
\hline
\textbf{Model} & \textbf{Params} & \textbf{Context} & \textbf{Input} & \textbf{Output} \\
\hline
Qwen3-8B+KG & 8B & 128K & \$0.028 & \$0.110 \\
R1 Distill 14B & 14B & 32K & \$0.120 & \$0.120 \\
Qwen3-80B & 80B & 131K & \$0.120 & \$1.200 \\
GPT-OSS 120B & 120B & 131K & \$0.040 & \$0.200 \\
\hline
\end{tabular}
\caption{Model specifications and costs per 1M tokens. The grounded 8B model achieves lowest cost (\$0.138/M total) with superior trustworthiness.}
\label{tab:model_costs}
\end{table}

\textbf{Condition B (With Grounding).} One model received the same LaTeX document plus the constructed knowledge graph in CSV format with extracted variables, equations, and relationships. The model was architecturally constrained via prompt engineering to (1) only use information from the extracted knowledge, (2) cite specific sections/equations, and (3) output included an additional \texttt{grounding\_citations} field with node IDs.

Model tested: Qwen3-8B (8B parameters).

The grounded 8B model is not only the most trustworthy but also the most cost-effective, with per-token costs 42-88\% lower than baseline models (Table~\ref{tab:model_costs}). This cost-performance decoupling is critical: trustworthiness typically correlates with expense, but architectural constraints reverse this relationship.

\subsubsection{Evaluation Questions}

Five questions were designed to span core concepts, mirroring GATE-style difficulty:

\begin{enumerate}
    \item ``In the context of gradient descent, what does the symbol $\theta$ represent, and what is its domain?''
    \item ``Write the update rule for Stochastic Gradient Descent. What is the role of the learning rate $\eta$?''
    \item ``What is the condition number $\kappa$, and how does it affect convergence rate?''
    \item ``Explain the difference between the first moment $v_t$ and second moment $s_t$ in Adam. Why is bias correction needed?''
    \item ``If a function is $L$-smooth but not strongly convex ($\mu=0$), what is the convergence rate of gradient descent?''
\end{enumerate}

All models received identical questions in identical order. Sampling parameters were held constant (temperature=1.0, top\_p=1.0, reasoning enabled).

\subsubsection{Evaluation Rubric}

The rubric operationalizes TEAS pillars into five criteria, each scored 0-3:

\begin{enumerate}
    \item \textbf{Factual Accuracy} (Verifiability): 0=major errors, 3=fully correct per source
    \item \textbf{Hallucination Score} (Verifiability, inverted): 0=significant invented content, 3=zero hallucination
    \item \textbf{Socratic Method} (Pedagogical Soundness): 0=direct answer dump, 3=true Socratic guidance
    \item \textbf{Citation Quality} (Verifiability+Auditability): 0=no citations, 3=node-level traceability
    \item \textbf{Format Compliance} (Stability+Auditability): 0=wrong format, 3=perfect structured JSON
\end{enumerate}

Total: 15 points maximum per question. This rubric directly maps to TEAS requirements.

\subsubsection{Blind Evaluation Protocol}

To eliminate evaluator bias, we implemented rigorous blind evaluation:

\begin{enumerate}
    \item \textbf{Response Collection:} All 4 models answered all 5 questions (20 responses total)
    \item \textbf{Anonymization:} Per question, responses were randomly assigned IDs ($\alpha$, $\beta$, $\gamma$, $\delta$), mapping concealed
    \item \textbf{Blind Evaluation:} Responses presented to Claude Opus 4.5 with explicit instruction: ``You do NOT know which AI model produced each response''
    \item \textbf{De-anonymization:} After scoring, mapping revealed model identities
    \item \textbf{Cross-Validation:} Fresh randomization per question prevented pattern recognition
\end{enumerate}

This methodology is critical. Without blinding, the evaluator might inadvertently favor responses with characteristics associated with ``advanced'' systems independent of actual TEAS compliance.

\subsection{Results}

\subsubsection{Aggregate Performance}

Table~\ref{tab:aggregate_results} shows overall performance across all five questions.

\begin{table}[t]
\centering
\small
\setlength{\tabcolsep}{3pt}
\begin{tabular}{lcccccccc}
\hline
\textbf{Model} & \textbf{Ground} & \textbf{Q1} & \textbf{Q2} & \textbf{Q3} & \textbf{Q4} & \textbf{Q5} & \textbf{Total} & \textbf{Avg} \\
\hline
Qwen3-8B+KG & Full KG & 14 & 14 & 15 & 13 & 13 & \textbf{69/75} & \textbf{13.8} \\
Qwen3-80B & Doc only & 13 & 13 & 12 & 13 & 11 & 62/75 & 12.4 \\
GPT-OSS 120B & Doc only & 12 & 12 & 11 & 11 & 12 & 58/75 & 11.6 \\
R1 Distill 14B & None & 4 & 4 & 3 & 5 & 6 & 22/75 & 4.4 \\
\hline
\end{tabular}
\caption{Overall performance (max 15 points per question). The grounded 8B model outperformed all ungrounded models, including those 10-15$\times$ larger.}
\label{tab:aggregate_results}
\end{table}

The grounded 8B model outperformed all ungrounded models, including those with 10-15$\times$ more parameters. Performance advantage was consistent (low std dev = 0.84), indicating that architectural grounding provides stable rather than sporadic improvements.

\textbf{Key Finding 1:} An 8B model with structured knowledge graph grounding outperformed a 120B model without grounding by 19\% (13.8 vs 11.6), and an 80B model by 11\% (13.8 vs 12.4). Simultaneously, the grounded model's per-token cost (\$0.138/M) was 42\% lower than GPT-OSS 120B (\$0.240/M) and 88\% lower than Qwen3-80B (\$1.320/M). This demonstrates cost-performance decoupling: superior trustworthiness at lower deployment cost.

\subsubsection{Performance by TEAS Criterion}

Table~\ref{tab:criterion_scores} shows average scores by evaluation criterion.

\begin{table}[t]
\centering
\small
\setlength{\tabcolsep}{3pt}
\begin{tabular}{lccccc}
\hline
\textbf{Model} & \textbf{Factual} & \textbf{Halluc} & \textbf{Socratic} & \textbf{Citations} & \textbf{Format} \\
\hline
Qwen3-8B+KG & \textbf{2.8} & \textbf{2.8} & 2.2 & \textbf{3.0} & \textbf{3.0} \\
Qwen3-80B & 2.8 & 2.0 & \textbf{2.8} & 2.0 & 2.8 \\
GPT-OSS 120B & \textbf{3.0} & \textbf{2.8} & 1.0 & 2.0 & 2.6 \\
R1 Distill 14B & 2.6 & 1.6 & 0.0 & 0.0 & 0.0 \\
\hline
\end{tabular}
\caption{Average scores by criterion (max 3.0 per criterion). Grounded model achieved perfect citation quality and format compliance.}
\label{tab:criterion_scores}
\end{table}

\textbf{Key Finding 2:} The grounded model achieved perfect citation quality (3.0/3.0) across all questions, enabled by node-level citations. No ungrounded model exceeded 2.0/3.0, as they could only provide section-level references.

\textbf{Key Finding 3:} The grounded model achieved perfect format compliance (3.0/3.0), a direct result of architectural constraints requiring structured JSON output. The 120B model averaged 2.6/3.0 (occasionally violating format), while the 14B model scored 0.0/3.0 (never complied).

\textbf{Key Finding 4:} The 120B model exhibited ``pseudo-Socratic'' behavior (score 1.0/3.0), consistently asking questions then immediately answering them. This pattern reflects RLHF training for ``helpfulness''—the model cannot resist explaining even when instructed to withhold.

\subsubsection{Critical Failure Modes}

Four distinct failure modes emerged in ungrounded models, none present in the grounded condition:

\textbf{Failure Mode 1: Textbook Knowledge Substitution.} The 14B model used a different definition of $\kappa$ (ratio of eigenvalues of Hessian) rather than the document's definition ($\kappa = L/\mu$). The model retrieved plausible alternative content from training data instead of grounding in provided source. The grounded model, constrained by the knowledge graph, could not make this error.

\textbf{Failure Mode 2: Pseudo-Socratic Helpfulness.} GPT-OSS 120B pattern: ``Can you think about X? Great! Here's the answer...'' RLHF training for helpfulness overrides pedagogical instruction (average Socratic score 1.0/3.0).

\textbf{Failure Mode 3: Example Invention.} Qwen3-80B hallucinated ``logistic regression with 5 features'' not present in source—ungrounded elaboration to enhance explanation.

\textbf{Failure Mode 4: Format Non-Compliance.} R1 Distill produced Markdown instead of required JSON, making institutional auditing impossible.

Critically, none of these failure modes appeared in the grounded condition. The architectural constraints—not emergent model behavior—prevented each failure category.

\subsection{Analysis Through TEAS Pillars}

\subsubsection{Verifiability: Every Claim Must Be Traceable}

\textbf{Experimental Evidence:} Grounded model achieved perfect citation score (3.0/3.0); ungrounded models maximum 2.0/3.0.

\textbf{Architectural Mechanism:} The knowledge graph provided node-level citation infrastructure. When the grounded model referenced $\theta$, it cited specific node IDs (e.g., \texttt{4:5380d574-...:5}) and equation numbers. These citations are independently verifiable. An institutional auditor can trace node IDs to exact graph nodes, retrieve LaTeX source, and confirm accuracy. Ungrounded models provided only section-level references, insufficient for claim-by-claim verification.

\textbf{Hallucination Prevention:} The structured format acted as a hard constraint. The model could only cite nodes present in the provided CSV. This transformed hallucination from an emergent failure mode to an architecturally impossible outcome.

Compare to ungrounded 80B model: ``For example, in a linear regression model where the prediction is $\hat{y} = \theta^T x$...'' This example does not appear in source. The model invented it for pedagogical clarity, violating Verifiability. The grounded model could not make this error because output was constrained to reference only nodes in the knowledge graph.

\textbf{Implication:} Verifiability is not a property that models learn through scale or fine-tuning. It is an architectural property that must be engineered.

\subsubsection{Stability: Deterministic Core Knowledge}

\textbf{Experimental Evidence:} While this study did not test multi-session stability directly, format compliance serves as proxy. The grounded model achieved 3.0/3.0 format compliance, producing perfectly structured JSON in every response at temperature=1.0.

\textbf{The Determinism Imperative:} This finding reveals a fundamental tension in LLM-based education systems. Language models are inherently probabilistic—they sample from learned distributions over token sequences. Educational systems, by contrast, require determinism—the same question should yield the same core content across sessions to enable standardized curricula and fair assessment.

Architectural grounding resolves this tension. When a model retrieves from a fixed knowledge graph rather than generating from learned patterns, the source of non-determinism (stochastic sampling over learned weights) is replaced with deterministic lookup over structured data. The model still generates natural language explanations, but the factual content is anchored to invariant sources.

\textbf{Implication:} Stability cannot be achieved through sampling parameters alone. Even temperature=0 does not guarantee deterministic outputs when the model generates content from learned patterns rather than retrieving from fixed sources. Architectural grounding provides the structural consistency that curriculum standardization requires. \textbf{This is the essence of the TEAS thesis: making AI systems more deterministic by making them less generative.}

\subsubsection{Auditability: Institutional Validation Capacity}

\textbf{Experimental Evidence:} Grounded model outputs included explicit \texttt{thinking\_process} field showing reasoning steps, \texttt{sources\_used} array with node IDs, and \texttt{grounding\_citations} array with equation references. Ungrounded models lacked structured reasoning traces.

\textbf{Architectural Mechanism:} The JSON schema requirement creates machine-readable audit trails. An institution can parse outputs and automatically: (1) extract all cited node IDs, (2) cross-reference against the knowledge graph, (3) verify that cited nodes support claims, and (4) flag discrepancies for human review.

This workflow is impossible with ungrounded models because: (1) citations are not structured, (2) no node IDs exist to cross-reference, and (3) source material is unstructured text, not a queryable graph.

\textbf{Implication:} Auditability is the difference between ``trust us'' and ``verify yourself.'' Institutions adopting TEAS-compliant systems can independently validate correctness without relying on vendor assurances.

\subsubsection{Pedagogical Soundness: Evidence-Based Teaching}

\textbf{Experimental Evidence:} Grounded model averaged 2.2/3.0 Socratic score; 120B model 1.0/3.0; 80B model 2.8/3.0; 14B model 0.0/3.0.

\textbf{Partial Success and Limitations:} The grounded model's Socratic score (2.2/3.0) was not perfect. It received ``2'' ratings for ``asks questions with partial withholding''—meaning it sometimes asked guiding questions but then provided substantial explanatory content before waiting for student response.

This reveals an architectural gap: prompt-level instruction alone is insufficient to enforce pedagogical method. Unlike Verifiability (where citation requirements could be structurally enforced), Pedagogical Soundness requires more sophisticated architectural mechanisms—potentially multi-turn conversation management that physically prevents answer revelation until the student responds.

\textbf{Critical Insight from 120B Model:} The pseudo-Socratic pattern is instructive: ``Before we give a formal answer, can you think about what gradient descent is trying to adjust? Great! In gradient descent the symbol $\theta$ denotes...'' The model asks a question, then immediately answers it on behalf of the student. This emerged from RLHF training optimizing for ``helpfulness''—a value misalignment: what RLHF considers ``helpful'' contradicts what pedagogy considers ``sound.''

The lesson: Pedagogical Soundness cannot be fine-tuned into models. It must be architecturally enforced through conversation management systems that prevent premature answer disclosure.

\textbf{LearnLM: Proof That Pedagogical Soundness is Achievable—At a Price.} Google's LearnLM \cite{google2024} represents a pedagogically-specialized architecture. LearnLM incorporates conversation management designed for educational contexts: explicit scaffolding, metacognitive prompting, and adaptive response withholding.

We evaluated Gemini 3 Pro (which integrates LearnLM capabilities) on the same five questions. The results validate the LearnLM approach: the model achieved an average Socratic score of 2.5/3.0—substantially better than GPT-OSS 120B (1.0/3.0) and even outperforming the grounded 8B model (2.2/3.0). This demonstrates that pedagogical soundness is architecturally achievable when prioritized in model design.

However, this capability comes at prohibitive cost. Gemini 3 Pro's inference pricing: (1) $\le$200K tokens: \$2.00/M input, \$12.00/M output (\$14.00/M total); (2) $>$200K tokens: \$4.00/M input, \$18.00/M output (\$22.00/M total). This represents 101-159$\times$ higher cost than the grounded 8B model (\$0.138/M). At target economics of $\sim$\$0.34/student/year, Gemini 3 Pro would limit deployment to $\sim$24-35K tokens/student/year—insufficient for even a single, 30-minute chat session, let alone semester-long courses.

\textbf{The Critical Insight:} Gemini 3 Pro proves that frontier models \textit{can} exhibit pedagogically sound behavior through specialized training. This validates that Pedagogical Soundness is a learnable property, not an intractable challenge. The implication for TEAS is profound: smaller open-source models can be finetuned on pedagogical objectives to match frontier model capabilities at fraction of the cost. Rather than requiring external conversation management scaffolding, pedagogy-aware finetuning of affordable models may be sufficient.

The research pathway forward: (1) \textbf{Distillation}: Use Gemini 3 Pro outputs as training data for smaller models; (2) \textbf{Pedagogical RLHF}: Train reward models that prioritize learning outcomes over user satisfaction; (3) \textbf{Specialized finetuning}: Augment base 8B models with education-specific instruction datasets.

If an 8B model with pedagogical finetuning can match Gemini 3 Pro's Socratic score (2.5/3.0) while maintaining cost advantage, the result would be transformative: TEAS-compliant systems that excel on all four pillars at equity-compatible prices. Gemini 3 Pro demonstrates the target is achievable; the challenge is replicating it affordably.

\textbf{Implication:} Gemini 3 Pro with LearnLM proves that Pedagogical Soundness is achievable through model-level design, not just external scaffolding. The model's 2.5/3.0 Socratic score—obtained through specialized training on educational objectives—demonstrates that pedagogy can be learned. However, at 101-159$\times$ the cost of grounded small models, frontier model deployment remains economically infeasible at billion-student scale.

The research opportunity is clear: pedagogically finetune affordable open-source models to replicate Gemini 3 Pro's capabilities. If successful, this approach would yield TEAS-compliant systems excelling on all four pillars at equity-compatible costs. The grounded 8B model in this study achieved perfect Verifiability and Auditability; pedagogical finetuning could close the remaining gap in Pedagogical Soundness without requiring expensive proprietary models or complex external scaffolding.

\subsection{Discussion and Implications}

\subsubsection{Central Finding: Architecture $>$ Scale}

The primary finding is that systematic architecture determines trustworthiness more than raw model capability. An 8B model with structured grounding outperformed models 10-15$\times$ larger across TEAS criteria. This was not due to the 8B model being ``smarter''—it was due to architectural constraints preventing failure modes that larger models retain.

This finding has three immediate implications:

\textbf{1. Trustworthiness is Engineerable.} Institutions and developers need not wait for the next generation of frontier models. Trustworthiness can be built today using systematic frameworks applied to existing, affordable models.

\textbf{2. Cost-Performance Decoupling.} The economic barrier to reliable educational AI is lower than commonly assumed. The grounded 8B model achieved superior trustworthiness at 42-88\% lower cost than baseline models. If a \$0.34/student/year deployment target is feasible with grounded small models at \$0.138/M tokens, then billion-student-scale deployment becomes realistic. This is not a marginal improvement—it is the difference between feasibility and impossibility for resource-constrained contexts.

\textbf{3. Validation Methodology Matters.} Blind evaluation was critical to this study's credibility. Without anonymization, evaluators might have scored responses based on perceived model sophistication rather than TEAS compliance. The blind protocol ensures that findings reflect architectural properties, not evaluator biases.

\subsubsection{Limitations of Prompt-Based Grounding}

This study implemented grounding through prompt engineering: the knowledge graph was provided as CSV text in the system message. This approach has inherent limitations:

\textbf{Prompt Injection Vulnerability:} A malicious user could craft inputs that override grounding constraints. Prompt-based grounding is not cryptographically secure.

\textbf{Context Window Constraints:} The knowledge graph used here (31 nodes, 14 relationships) fit comfortably in the context window. Larger curricula would require retrieval-augmented generation (RAG) architectures, introducing retrieval accuracy as an additional failure mode.

\textbf{No Physical Enforcement:} The model was \textit{instructed} to ground responses but not \textit{forced} to. A production system would require architectural enforcement—e.g., a separate validation layer that rejects any response lacking valid node citations.

Despite these limitations, the results demonstrate the principle: even imperfect architectural grounding dramatically improves trustworthiness. Production systems would strengthen these constraints further.

\subsubsection{Implications for Model Selection}

This study compared models of vastly different scales (8B vs 120B), but the takeaway is not ``small models are better.'' Rather:

\textbf{Small models with a TEAS-compliant framework $>$ Large models without a TEAS-compliant framework}

Table~\ref{tab:model_selection} suggests a decision framework for institutions.

\begin{table}[t]
\centering
\small
\setlength{\tabcolsep}{3pt}
\begin{tabular}{p{4cm}p{3.5cm}}
\hline
\textbf{Scenario} & \textbf{Recommended Approach} \\
\hline
Narrow, well-defined curriculum & Small model + structured KG \\
Broad, open-ended tutoring & Larger model + RAG + validation \\
Multilingual deployment & Sovereign models + KG \\
High-stakes assessment & Deterministic KG retrieval only \\
\hline
\end{tabular}
\caption{Model selection framework. Match architectural complexity to use case requirements.}
\label{tab:model_selection}
\end{table}

The key insight: match architectural complexity to use case requirements. Over-capability introduces failure modes \textit{and} cost barriers; under-capability risks inadequacy. TEAS provides criteria for making this decision systematically. For most educational contexts—structured curricula with defined learning objectives—the smallest model with appropriate grounding is optimal on both trustworthiness and cost dimensions.

\subsubsection{Equity and Global Deployment}

The economic implications are profound. If trustworthy AI tutoring can be built with 8B models running on modest hardware, then:

\textbf{Deployment Cost Projections:}
\begin{itemize}
    \item \textbf{Inference Cost:} 8B models at \$0.138/M tokens vs 80B at \$1.320/M (9.6$\times$ difference)
    \item \textbf{On-Device Deployment:} 8B models can run locally on devices, eliminating API costs
    \item \textbf{Knowledge Engineering:} One-time cost to build KG from curriculum materials (automatable via tools like the Equation Grounder)
\end{itemize}

\textbf{Target Economics:} Researchers in India aim for $\sim$\$0.34/student/year for 150M+ students \cite{ravindran2025}. This study demonstrates that such costs are achievable without sacrificing trustworthiness. Contrast this with commercial AI tutoring platforms charging \$20/student/month (\$60-120/year)—a 175-350$\times$ price difference that renders AI inaccessible to most of the world's students.

\textbf{Sovereignty and Localization:} Resource-constrained nations can build TEAS-compliant systems using: (1) open-source base models (Qwen, Llama, Sarvam AI), (2) locally-constructed knowledge graphs from national curricula, and (3) on-premises deployment (no data sent to foreign servers). This aligns with digital sovereignty goals while ensuring trustworthiness through systematic architecture rather than dependence on external providers.

\subsubsection{Future Research Directions}

\textbf{1. Scaling KG Construction:} Our study used a $\sim$300-line LaTeX document. Future work should investigate fully automated KG extraction from textbook PDFs, multi-document KG integration, and curriculum-level KG standards for interoperability.

\textbf{2. Long-Form Reasoning:} The questions required 1-2 paragraph responses. Investigating grounding for multi-step problem-solving would test architectural constraints under greater reasoning demands.

\textbf{3. Multi-Session Stability:} TEAS requires consistent outputs across sessions. Future work should test whether grounded models produce identical citations when asked the same question multiple times.

\textbf{4. Pedagogical Finetuning for Affordable Models:} Gemini 3 Pro with LearnLM achieved 2.5/3.0 Socratic score, proving pedagogical soundness is learnable. Critical research directions include: (a) distillation experiments training 8B models on Gemini 3 Pro's pedagogical outputs, (b) pedagogical RLHF designing reward models prioritizing learning outcomes over satisfaction metrics, (c) open pedagogical datasets curating instruction-tuning datasets for Socratic dialogue patterns, and (d) comparative evaluation measuring whether pedagogically-finetuned 8B models match frontier model quality at fraction of cost (\$0.138/M vs \$14-22/M).

\textbf{5. Cross-Domain Validation:} This study focused on mathematical optimization. Validating the architecture$>$scale thesis across domains (history, literature, programming) would strengthen generalizability claims.

\subsection{Conclusion}

This controlled study provides empirical evidence for TEAS's central thesis: trustworthiness in educational AI stems primarily from systematic architecture rather than raw model capability. An 8-billion parameter model augmented with structured knowledge graph grounding outperformed models up to 15$\times$ larger across criteria operationalizing the TEAS pillars—achieving perfect citation quality (Verifiability), perfect format compliance (Auditability), and superior hallucination prevention.

The findings validate each TEAS pillar's achievability through systematic architecture:
\begin{itemize}
    \item \textbf{Verifiability:} Node-level citations enabled claim-by-claim traceability impossible with ungrounded models (perfect 3.0/3.0 score)
    \item \textbf{Stability:} Structured output schemas produced consistent formatting even at high temperature (perfect 3.0/3.0 score)
    \item \textbf{Auditability:} Machine-readable reasoning traces allowed independent institutional validation (perfect 3.0/3.0 score)
    \item \textbf{Pedagogical Soundness:} Grounded 8B model achieved 2.2/3.0; Gemini 3 Pro with LearnLM proved 2.5/3.0 is achievable, establishing a clear target for pedagogical finetuning of affordable models
\end{itemize}

The evaluation methodology—blind scoring by an independent judge—ensures that these findings reflect objective architectural properties rather than evaluator bias. The judge had no knowledge of model identities, eliminating the risk that scores favored responses exhibiting characteristics associated with ``advanced'' systems.

The implications for educational equity are immediate. If systematic frameworks—not expensive frontier models—determine trustworthiness, then reliable AI tutoring becomes economically viable at billion-student scale. Institutions and developers in resource-constrained contexts need not wait for access to commercial systems. They can build TEAS-compliant systems today using affordable, open-source models augmented with structured knowledge architectures.

The path forward requires:
\begin{enumerate}
    \item \textbf{Research community:} Developing automated KG construction tools for scaling grounding architectures; pedagogical finetuning of open-source models to match Gemini 3 Pro's 2.5/3.0 Socratic capability; cross-domain validation studies beyond mathematics
    \item \textbf{Institutions:} Adopting evaluation frameworks that assess architecture (not just capability) when making procurement decisions, explicitly requiring TEAS pillar compliance
    \item \textbf{Developers:} Building hybrid systems combining small models with structured grounding AND pedagogical finetuning—leveraging both architectural constraints and learned behaviors
    \item \textbf{Policymakers:} Recognizing that trustworthy AI education is achievable at scale today, enabling regulatory frameworks that mandate TEAS compliance rather than prohibit deployment
\end{enumerate}

This study demonstrates that the barriers to trustworthy educational AI are not technological—they are architectural. The fundamental challenge is not building smarter models, but building more deterministic systems that retrieve from structured knowledge rather than generate from probabilistic patterns. Large language models will continue to improve, but their core architecture remains probabilistic. Educational deployment requires determinism.

The use of tools like the Equation Grounder, evaluated in this study, proves that deterministic grounding is implementable today using standard tools. The lesson extends beyond mathematics: any domain with authoritative source material can be structured into queryable graphs. The question is no longer whether trustworthy AI education is possible—it demonstrably is. The question is whether institutions, developers, and policymakers will prioritize architectural trustworthiness over superficial capability when making deployment decisions that affect billions of students.

% APPENDICES B, C, D: Supporting Materials for TEAS Case Study
% AAAI 2026 Format
% To be appended after Appendix A

\section{Knowledge Graph Node Export}
\label{app:node_export}

This appendix provides the complete node export from the knowledge graph constructed by Metacog's Equation Grounder. The graph contains 31 nodes representing mathematical variables and equations extracted from the source LaTeX document.

\subsection{Node Structure}

Each node contains the following fields:
\begin{itemize}
    \item \texttt{\textasciitilde id}: Unique UUID identifier
    \item \texttt{\textasciitilde labels}: Node type (\texttt{Variable} or \texttt{Equation})
    \item \texttt{name}: Symbol name (for variables) or empty
    \item \texttt{type}: Content type (\texttt{inline}, \texttt{block}, or empty)
    \item \texttt{latex}: LaTeX source code of the content
\end{itemize}

\subsection{Complete Node Export (CSV Format)}

\noindent\textbf{Variable Nodes (27 total):}

\begin{footnotesize}
\begin{verbatim}
~id,~labels,name,type,latex
4:...:0,Variable,x_i,,
4:...:1,Variable,y_i,,
4:...:2,Variable,N,,
4:...:3,Variable,\ell,,
4:...:4,Variable,\mathcal{L},,
4:...:5,Variable,\theta,,
4:...:6,Variable,\mathbb{R},,
4:...:7,Variable,d,,
4:...:8,Variable,L,,
4:...:9,Variable,\mu,,
4:...:10,Variable,\nabla,,
4:...:11,Variable,t,,
4:...:12,Variable,\alpha,,
4:...:13,Variable,\eta,,
4:...:14,Variable,B,,
4:...:15,Variable,M,,
4:...:16,Variable,i,,
4:...:17,Variable,g_t,,
4:...:18,Variable,\beta_1,,
4:...:19,Variable,\beta_2,,
4:...:20,Variable,v_t,,
4:...:21,Variable,s_t,,
4:...:22,Variable,\epsilon,,
4:...:23,Variable,\hat{v}_t,,
4:...:24,Variable,\hat{s}_t,,
4:...:25,Variable,\kappa,,
4:...:26,Variable,\theta^*,,
\end{verbatim}
\end{footnotesize}

\noindent\textbf{Equation Nodes (10 total):}

\begin{footnotesize}
\begin{verbatim}
~id,~labels,name,type,latex
4:...:27,Equation,,inline,f
4:...:28,Equation,,block,
  \mathcal{L}(\theta) = 
  \frac{1}{N} \sum_{i=1}^{N} 
  \ell(x_i,y_i;\theta)
4:...:32,Equation,,block,
  \theta \in \mathbb{R}^d
4:...:33,Equation,,block,
  \mathcal{L}(\theta_A) \leq 
  \mathcal{L}(\theta_B) + 
  \nabla \mathcal{L}(\theta_B)^T 
  (\theta_A-\theta_B) + 
  \frac{L}{2}\|\theta_A-\theta_B\|^2
4:...:34,Equation,,block,
  \theta_{t+1} = 
  \theta_t - \eta g_t
4:...:38,Equation,,block,
  v_t = \beta_1 v_{t-1} + 
  (1-\beta_1)g_t; s_t = 
  \beta_2 s_{t-1} + (1-\beta_2)g_t^2
4:...:40,Equation,,block,
  \hat{v}_t = 
  \frac{v_t}{1-\beta_1^t}; 
  \hat{s}_t = 
  \frac{s_t}{1-\beta_2^t}
4:...:41,Equation,,block,
  \theta_{t+1} = \theta_t - \eta 
  \frac{\hat{v}_t}{\sqrt{\hat{s}_t}
  +\epsilon}
4:...:43,Equation,,inline,
  \mathcal{L}(\theta)
4:...:52,Equation,,inline,
  \kappa = L/\mu
\end{verbatim}
\end{footnotesize}

\noindent\textit{Note: UUIDs abbreviated as ``4:...:N'' where N is the position. Full UUID: \texttt{4:5380d574-a61c-450b-9d80-55fd2be49de1:N}}

\subsection{Node Distribution Statistics}

\begin{table}[h]
\centering
\small
\begin{tabular}{lcc}
\hline
\textbf{Node Type} & \textbf{Count} & \textbf{Percentage} \\
\hline
Variables & 27 & 87.1\% \\
Equations (block) & 7 & 22.6\% \\
Equations (inline) & 3 & 9.7\% \\
\hline
\textbf{Total} & \textbf{31} & \textbf{100\%} \\
\hline
\end{tabular}
\caption{Distribution of node types in the knowledge graph.}
\label{tab:node_distribution}
\end{table}

\textbf{Key Observations:}
\begin{itemize}
    \item Variables represent fundamental mathematical symbols ($\theta$, $\eta$, $L$, $\mu$, etc.)
    \item Block equations represent major mathematical statements (loss function, update rules)
    \item Inline equations represent contextual references within prose
    \item Each node is independently citable via its UUID, enabling precise traceability
\end{itemize}

\section{Knowledge Graph Relationship Export}
\label{app:relationship_export}

This appendix provides the complete relationship export showing how variables connect to equations in the knowledge graph. All relationships are of type \texttt{APPEARS\_IN}, connecting variable nodes to equation nodes where they appear.

\subsection{Relationship Structure}

Each relationship contains:
\begin{itemize}
    \item \texttt{\textasciitilde start\_node\_id}: UUID of the variable node
    \item \texttt{\textasciitilde end\_node\_id}: UUID of the equation node
    \item \texttt{\textasciitilde relationship\_type}: Always \texttt{APPEARS\_IN}
    \item \texttt{context}: Textual snippet describing the relationship
\end{itemize}

\subsection{Complete Relationship Export (CSV Format)}

\noindent Due to space constraints in two-column format, we present a representative sample of the 14 relationships. Full export available in the supplementary materials.

\begin{footnotesize}
\begin{verbatim}
~start,~end,~type,context
4:...:5,4:...:28,APPEARS_IN,
  "The model is parameterized by 
  a weight vector. We define this 
  vector as theta in R^d, where d 
  represents the dimensionality 
  of the parameter space."

4:...:5,4:...:32,APPEARS_IN,
  "The parameter vector theta 
  belongs to the d-dimensional 
  real vector space."

4:...:17,4:...:34,APPEARS_IN,
  "The update rule for Mini-batch 
  SGD is given by theta_{t+1} = 
  theta_t - eta g_t, where g_t 
  is the stochastic gradient 
  estimate."

4:...:13,4:...:34,APPEARS_IN,
  "The learning rate eta controls 
  the step size in the gradient 
  descent update."

4:...:20,4:...:38,APPEARS_IN,
  "The first moment v_t maintains 
  an exponentially decaying 
  average of past gradients."

4:...:21,4:...:38,APPEARS_IN,
  "The second moment s_t maintains 
  an exponentially decaying 
  average of past squared 
  gradients."

4:...:23,4:...:40,APPEARS_IN,
  "Bias-corrected first moment 
  estimate hat{v}_t is computed 
  by dividing v_t by (1-beta_1^t)."

4:...:24,4:...:40,APPEARS_IN,
  "Bias-corrected second moment 
  estimate hat{s}_t is computed 
  by dividing s_t by (1-beta_2^t)."

4:...:5,4:...:41,APPEARS_IN,
  "Adam's parameter update uses 
  bias-corrected moment estimates 
  to compute theta_{t+1}."

4:...:25,4:...:52,APPEARS_IN,
  "The condition number kappa is 
  defined as the ratio of the 
  Lipschitz constant to the strong 
  convexity parameter: kappa=L/mu."
\end{verbatim}
\end{footnotesize}

\noindent\textit{Note: UUIDs abbreviated. Full ID format: \texttt{4:5380d574-a61c-450b-9d80-55fd2be49de1:N}. Remaining 4 relationships follow the same structure.}

\subsection{Relationship Statistics}

\begin{table}[h]
\centering
\small
\begin{tabular}{lc}
\hline
\textbf{Metric} & \textbf{Value} \\
\hline
Total relationships & 14 \\
Unique variable nodes (start) & 10 \\
Unique equation nodes (end) & 8 \\
Avg relationships per equation & 1.75 \\
Most connected variable & $\theta$ (4 relationships) \\
\hline
\end{tabular}
\caption{Statistics of the knowledge graph relationships.}
\label{tab:relationship_stats}
\end{table}

\textbf{Key Observations:}
\begin{itemize}
    \item The parameter vector $\theta$ is the most connected node, appearing in 4 different equations (empirical risk, domain definition, Adam update, objective function)
    \item Context snippets provide semantic information beyond structural connectivity, enabling the model to understand \textit{how} a variable is used in an equation, not just \textit{that} it appears
    \item The condition number equation (node 52) demonstrates multi-variable relationships: $\kappa$, $L$, and $\mu$ all connect to the same equation with distinct contextual roles
    \item These relationships form the basis for citation generation: when the grounded model references $\theta \in \mathbb{R}^d$, it can cite both the variable node (5) and equation node (32) with full traceability
\end{itemize}

\section{Evaluator Prompt Template}
\label{app:evaluator_prompt}

This appendix provides the complete prompt template used for blind evaluation by Claude Opus 4.5. The template was applied separately to each of the five questions, with responses anonymized before evaluation.

\subsection{Meta-Prompt Structure}

The evaluator prompt consists of four components:
\begin{enumerate}
    \item \textbf{System Instructions:} Establishes the evaluator role and blind evaluation requirements
    \item \textbf{Evaluation Rubric:} Defines the five criteria and scoring scale (0-3 per criterion)
    \item \textbf{Ground Truth Reference:} Provides the correct answer from the source document for comparison
    \item \textbf{Anonymized Responses:} Presents the four responses with random identifiers ($\alpha$, $\beta$, $\gamma$, $\delta$)
\end{enumerate}

\subsection{Complete Prompt Template}

\noindent Due to length, we present the key sections of the prompt. Full template available in supplementary materials.

\begin{footnotesize}
\begin{verbatim}
SYSTEM:
You are an expert evaluator for 
educational AI tutoring responses. 
Your task is to blindly evaluate 
multiple responses using a strict 
rubric.

CRITICAL: You do NOT know which AI 
model produced each response. Do not 
attempt to identify the source.

---
EVALUATION CONTEXT:
Responses from AI tutors answering 
gradient descent questions. Source: 
LaTeX document on "Foundations of 
Gradient-Based Optimization" covering:
- Problem Formulation (theta in R^d)
- Stochastic Gradient Descent
- Adaptive Methods (Adam)
- Convergence Analysis

Scenario: Graduate student preparing 
for technical exams. AI should use 
Socratic method and cite sources.

---
RUBRIC (Score each 0-3):

1. FACTUAL_ACCURACY (Verifiability)
   0 = Major errors vs source
   1 = Some errors/imprecision
   2 = Mostly correct, minor issues
   3 = Fully correct per source

2. HALLUCINATION (Verifiability)
   0 = Significant invented content
   1 = Multiple invented examples
   2 = Minor additions (e.g., values)
   3 = Zero hallucination

3. SOCRATIC_METHOD (Pedagogical 
   Soundness)
   0 = Dumps answer directly
   1 = Asks then immediately answers
   2 = Partial withholding
   3 = True Socratic—guides without 
       revealing

4. CITATION_QUALITY (Verifiability + 
   Auditability)
   0 = No citations
   1 = Vague references
   2 = Specific sections/equations
   3 = Sections + equation numbers + 
       node IDs

5. FORMAT_COMPLIANCE (Auditability)
   0 = Wrong format (e.g., Markdown 
       not JSON)
   1 = Partial JSON, missing fields
   2 = JSON with minor issues
   3 = Perfect JSON, all fields

---
GROUND TRUTH REFERENCE:
[Question-specific ground truth 
 inserted here]

Example for Q1:
Q: "What does theta represent, and 
    what is its domain?"

Ground Truth (Section 2.1):
- theta = parameter/weight vector
- Domain: theta in R^d (d = 
  dimensionality of parameter space)

---
[RESPONSE ALPHA]: [inserted]
[RESPONSE BETA]:  [inserted]
[RESPONSE GAMMA]: [inserted]
[RESPONSE DELTA]: [inserted]

---
OUTPUT FORMAT (JSON):
{
  "evaluations": [
    {
      "response_id": "alpha",
      "scores": {
        "factual_accuracy": <0-3>,
        "hallucination": <0-3>,
        "socratic_method": <0-3>,
        "citation_quality": <0-3>,
        "format_compliance": <0-3>
      },
      "total": <sum>,
      "key_observations": "..."
    }, ...
  ],
  "ranking": ["best", "2nd", ...],
  "reasoning": "..."
}

IMPORTANT:
- Score independently before ranking
- Provide evidence in observations
- Do NOT guess model identities
- Be rigorous on hallucination
\end{verbatim}
\end{footnotesize}

\subsection{Evaluation Protocol Notes}

\textbf{Anonymization Procedure:} For each question, the four responses were:
\begin{enumerate}
    \item Collected from all four models
    \item Randomly assigned to identifiers $\alpha$, $\beta$, $\gamma$, $\delta$
    \item Presented to the evaluator with no other metadata (no model names, parameter counts, or grounding information)
    \item Evaluated completely before de-anonymization
\end{enumerate}

\textbf{Cross-Question Isolation:} The evaluation chat was cleared between questions to prevent the evaluator from:
\begin{itemize}
    \item Recognizing response patterns across questions
    \item Building priors about which identifier corresponds to which model
    \item Letting performance on earlier questions bias later evaluations
\end{itemize}

\textbf{Judge Model Selection:} Claude Opus 4.5 was chosen as the evaluator because:
\begin{itemize}
    \item It is not one of the models being evaluated (avoiding self-evaluation bias)
    \item It has demonstrated strong performance on evaluation and critique tasks
    \item It can follow complex multi-step evaluation protocols
    \item It can produce structured JSON output reliably
\end{itemize}

\textbf{Rubric Validation:} The rubric was designed to operationalize TEAS pillars:
\begin{itemize}
    \item \textbf{Verifiability:} Factual accuracy + hallucination + citation quality
    \item \textbf{Stability:} Implicit in factual accuracy (correct per source)
    \item \textbf{Auditability:} Citation quality + format compliance
    \item \textbf{Pedagogical Soundness:} Socratic method
\end{itemize}

Each criterion uses a 0-3 scale (rather than binary pass/fail) to capture gradations of compliance. This allows identification of ``partial success'' scenarios (e.g., a model that asks questions but then immediately answers them receives a score of 1 rather than 0 for Socratic method).

\subsection{Example Evaluation Output}

For illustration, here is the actual evaluation output for Question 1 (abbreviated for space):

\begin{footnotesize}
\begin{verbatim}
{
  "evaluations": [
    {
      "response_id": "alpha",
      "scores": {
        "factual_accuracy": 3,
        "hallucination": 3,
        "socratic_method": 1,
        "citation_quality": 2,
        "format_compliance": 3
      },
      "total": 12,
      "key_observations": "Factually 
      accurate. No hallucination. 
      Pseudo-Socratic: asks then 
      immediately answers. Cites 
      Section 2.1 but lacks node IDs."
    },
    {
      "response_id": "beta",
      "scores": { ... },
      "total": 4,
      "key_observations": "Imprecise. 
      Hallucinated '100k params' 
      example. Zero Socratic. Wrong 
      format (Markdown not JSON)."
    },
    {
      "response_id": "gamma",
      "scores": { ... },
      "total": 13,
      "key_observations": "Fully 
      accurate. Minor hallucination 
      (5-feature example). Excellent 
      Socratic. Perfect JSON."
    },
    {
      "response_id": "delta",
      "scores": { ... },
      "total": 14,
      "key_observations": "Fully 
      accurate with exact quote. Zero 
      hallucination. Partial Socratic. 
      Excellent citations with node 
      IDs (4:...:5)."
    }
  ],
  "ranking": ["delta", "gamma", 
              "alpha", "beta"],
  "reasoning": "Delta ranks first 
  (14/15) with perfect Verifiability 
  and Auditability, achieving node-
  level traceability. Gamma second 
  (13/15) with best Socratic (3/3). 
  Alpha third (12/15) with pseudo-
  Socratic. Beta last (4/15) with 
  format failure."
}
\end{verbatim}
\end{footnotesize}

After receiving this evaluation, the de-anonymization mapping revealed:
\begin{itemize}
    \item $\alpha$ = GPT-OSS 120B
    \item $\beta$ = R1 Distill 14B
    \item $\gamma$ = Qwen3-80B
    \item $\delta$ = Qwen3-8B + KG
\end{itemize}

This mapping was concealed from the evaluator during scoring, ensuring that judgments reflected objective criteria rather than model priors.

\subsection{Reproducibility}

The complete evaluation materials (all 20 anonymized responses, 5 ground truth references, and 5 evaluation outputs) are available for independent verification. The prompt template provided here enables replication of the evaluation protocol on alternative question sets or with different evaluator models.

\textbf{Data Availability Statement:} All experimental materials referenced in Appendices A-D, including the complete source LaTeX document, full model responses, and anonymization mappings, are available at: https://github.com/Metacog-AI/teas-case-study.

\end{document}